\documentclass{article}
\usepackage[T1]{fontenc}
\usepackage{setspace}
\usepackage[utf8]{inputenc}
\usepackage[letterpaper, total={6in, 9in}]{geometry}
\usepackage{color}
\usepackage{graphicx}
\usepackage{authblk}
\usepackage[bottom]{footmisc} % fix footnote

\usepackage[backend=biber,style=nature]{biblatex}
\title{\textbf{High resolution photonic force microscopy based on sharp nano-fabricated tips}}

\author[1,4]{Rudy Desgarceaux*}
\author[1]{Zhanna Santybayeva*}
\author[1]{Eliana Battistella}
\author[1]{Ashley L. Nord}
\author[2]{Catherine Braun-Breton}
\author[1]{Manouk Abkarian}
\author[3]{Onofrio M. Maragò}
\author[4]{Benoit Charlot$^{\dagger}$}
\author[1]{Francesco Pedaci$^{\dagger}$}

\affil[1]{\small{CBS Un.Montpellier, CNRS, INSERM, Montpellier, France}}
\affil[2]{Un.Montpellier, CNRS UMR
5235, Montpellier, France}
\affil[3]{\small{CNR-IPCF, Istituto per i Processi Chimico-Fisici, Messina, Italy}}
\affil[4]{\small{IES, CNRS UMR 5214, University of Montpellier, Montpellier, France}}
\affil[*]{\small{equal contribution}}
\affil[$\dagger$]{\small{corresponding authors}}
\date{}   

\begin{document}

\maketitle

\textbf{Sub-nm resolution images can be achieved by Atomic Force Microscopy (AFM) on samples that are deposited on hard substrates. However, it is still extremely challenging to image soft interfaces, such as biological membranes, due to the deformations induced by the tip. 
Photonic Force Microscopy (PhFM), based on optical tweezers (OT), represents an interesting alternative for soft scanning-probe microscopy \cite{ghislain1993scanning, florin1997photonic, friese1999three, tischer2001three, marago2010photonic, friedrich2015surface, jones2015optical,  bartsch2016nanoscopic}. 
Using light instead of a physical cantilever to hold the scanning probe results in a stiffness ($k_{OT}\sim0.1-0.001$ pN/nm) which can be 2-3 orders of magnitude lower than that of standard cantilevers ($k_{AFM}\sim 10$ pN/nm).
Combined with nm resolution of displacement measurements of the trapped probe, this allows for imaging soft materials without force-induced artefacts. 
However, the size of the optically trapped probe, often chosen as a $\sim \mu$m-size sphere, has so far limited the resolution of PhFM.
Here we show a novel and simple nanofabrication protocol to massively produce optically trappable quartz particles which mimic the sharp tips of AFM.
We demonstrate and quantify the stable trapping of particles with tips as sharp as 35 nm, the smallest used in PhFM to date. Raster scan images of rigid nanostructures with features smaller than 80 nm obtained with our tips compare well with AFM images of the same samples. Imaging the membrane of living malaria-infected red blood cells produces no visible artefacts and reveals the sub-micron structural features termed knobs, related to the parasite activity within the cell. The use of nano-engineered particles in PhFM opens the way to imaging soft and biological samples at high resolution.\\}

In PhFM, the laser trap raster scans the sample and the induced displacement of the trapped particle from its equilibrium position is measured by fast interferometry.
Other than providing sub-pN/nm stiffness, optically trapping the probe decouples it from the instrument, potentially allowing low-force scanning within confined volumes, not accessible by other techniques \cite{rohrbach2004trapping}. 
As in AFM, measurements of the axial displacement of the probe in PhFM (signal $S_z$) reach the nanometer resolution. 
Recently, an improved version of the PhFM \cite{friedrich2015surface} has achieved resolutions beyond the diffraction limit in the transverse directions (signals $S_x$ and $S_y$) using $\sim 100-200$ nm beads as scanning probes. 
When trapping a spherical particle, reducing its size   unfortunately reduces also the signal to noise ratio obtained, therefore other strategies are required to increase the lateral resolution.
In order to control the applied optical force and to increase the resolution, different particle geometries have been used in different configurations \cite{phillips2012optically, simpson2013bespoke, phillips2014shape, phillips2011surface, olof2012measuring}. 
In particular, the behavior of elongated particles trapped in OT has been studied experimentally and theoretically due to their potential use as scanning probes  \cite{kress2004tilt, nakayama2007tunable, bui2013calibration, wang2013resolving, griesshammer20145d}. 
However, the resolution of PhFM still remains limited to hundreds of nm in two of the three dimensions, and the PhFM lags behind the AFM mainly because sharp tips equivalent to those used in AFM have yet to be developed.

\begin{figure}[t]
\centering
\includegraphics[width=0.99\textwidth]{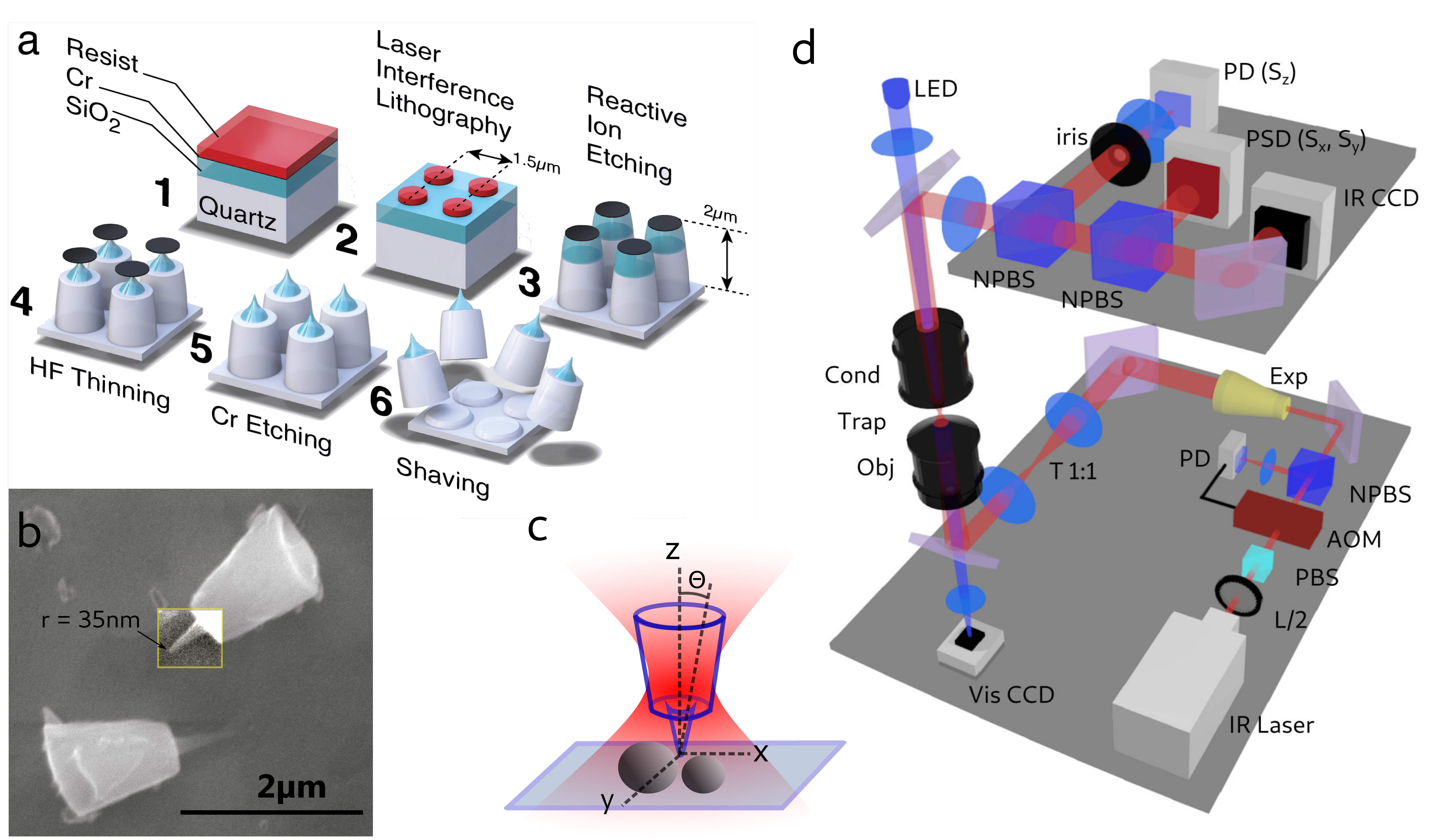}
\caption{Tip fabrication and optical setup. 
\textbf{a)} Microfabrications process. Cylindrical particles are generated by Laser Interference Lithography, etching a quartz substrate where a 800 nm thick SiO$_2$ layer is deposited. A tuned acid thinning by HF produces sharp tips in the SiO$_2$ layer. The particles are then cleaved mechanically off the substrate.
\textbf{b)} Scanning electron microscope image of cleaved particles, where the contrast of one tip has been enhanced for clarity. The radius of curvature of the tip is 35 nm.
\textbf{c)} Schematic cartoon of the optical trap holding the particle and scanning the surface of the sample with the sharp tip.
\textbf{d)} Schematic optical setup. L/2: half-wave plate, PBS: polarizer, AOM: acousto-optical modulator, NPBS: non-polarizing beam splitter, Exp: beam expander, T1:1 : one to one telescope, Obj: Objective, Cond: Condenser, PD: photodiode (to acquire $S_z$), PSD: position sensitive detector (to acquire $S_{x,y}$), IRCCD: infra red CCD camera, VisCCD: visible CCD camera.)}
\label{fig_1}
\end{figure}

We have developed a novel nanofabrication protocol (sketched in fig.\ref{fig_1}a) wherein we can produce optically trappable particles ($60\times 10^6$ per batch) which mimic the sharp tips of AFM. 
The simplicity of the nanofabrication protocol relies mainly on the use of Laser Interference Lithography (LIL) \cite{van2011laser} (a cost-effective, parallel and reliable technique which we describe in detail in \cite{santybayeva2016fabrication}) to write the pattern that produces the particles. 
To maximize the optical signals from the OT, the probe is designed as a micron-sized truncated cone holding a sharp feature (of 35 nm radius of curvature in fig.\ref{fig_1}b) in the center of its top surface. 
This geometry is produced by two serial etching processes which follow the LIL exposure: a first anisotropic Reactive Ion Etching to produce the cylindrical geometry, followed by an isotropic hydrofluoric acid (HF) wet etch step to produce the tips on a top deposited SiO$_2$ layer. 
Once obtained, the particles are mechanically cleaved off the substrate, collected, and stored in liquid. By SEM imaging, we find that 30\% of the collected particles maintain their tip.
The details of the nanofabrication protocol are described in the Methods section.

The core material chosen is quartz, a crystal whose birefringence is instrumental in PhFM.
Birefringent cylindrical particles have been developed by different techniques in the context of optical single-molecule manipulation \cite{deufel2007nanofabricated, gutierrez2010optical, huang2011electron, li2014fabrication, santybayeva2016fabrication, ha2019single}.
Due to the torque imposed by radiation pressure, the main axis of the trapped particle remains parallel to the laser propagation direction. 
Also, due to the torque imposed by the laser linear polarization on the birefringent medium, the x-cut quartz particle cannot rotate about its geometrical axis (the particle can rotate about its main axis if the laser polarization is circular or if the linear polarization is set in rotation) \cite{la2004optical, pedaci2011excitable}. 
This constrains all the degrees of freedom ($x,y,z$ and two angles, fig.\ref{fig_1}c) of the trapped particle.
%We will address below how we can recognize whether the tip points downwards (facing the sample) or upwards.
Our OT setup is shown in fig.\ref{fig_1}d, and is described in detail in the Methods section. Briefly, a stabilized infrared laser beam of 20-60 mW power is focused by a 1.2NA microscope objective into a flow cell containing the sample to scan. The forward-scattering detection consists of a position sensitive detector (PSD) to acquire the transverse displacement of the trapped particle (signals $S_x$ and $S_y$), and of a separate photodiode with reduced numerical aperture \cite{dreyer2004improved}, to detect the axial displacement ($S_z$ signal). In this way the transverse and axial resolution can be maximized independently.

\begin{figure}
\centering
\includegraphics[width=0.95\textwidth]{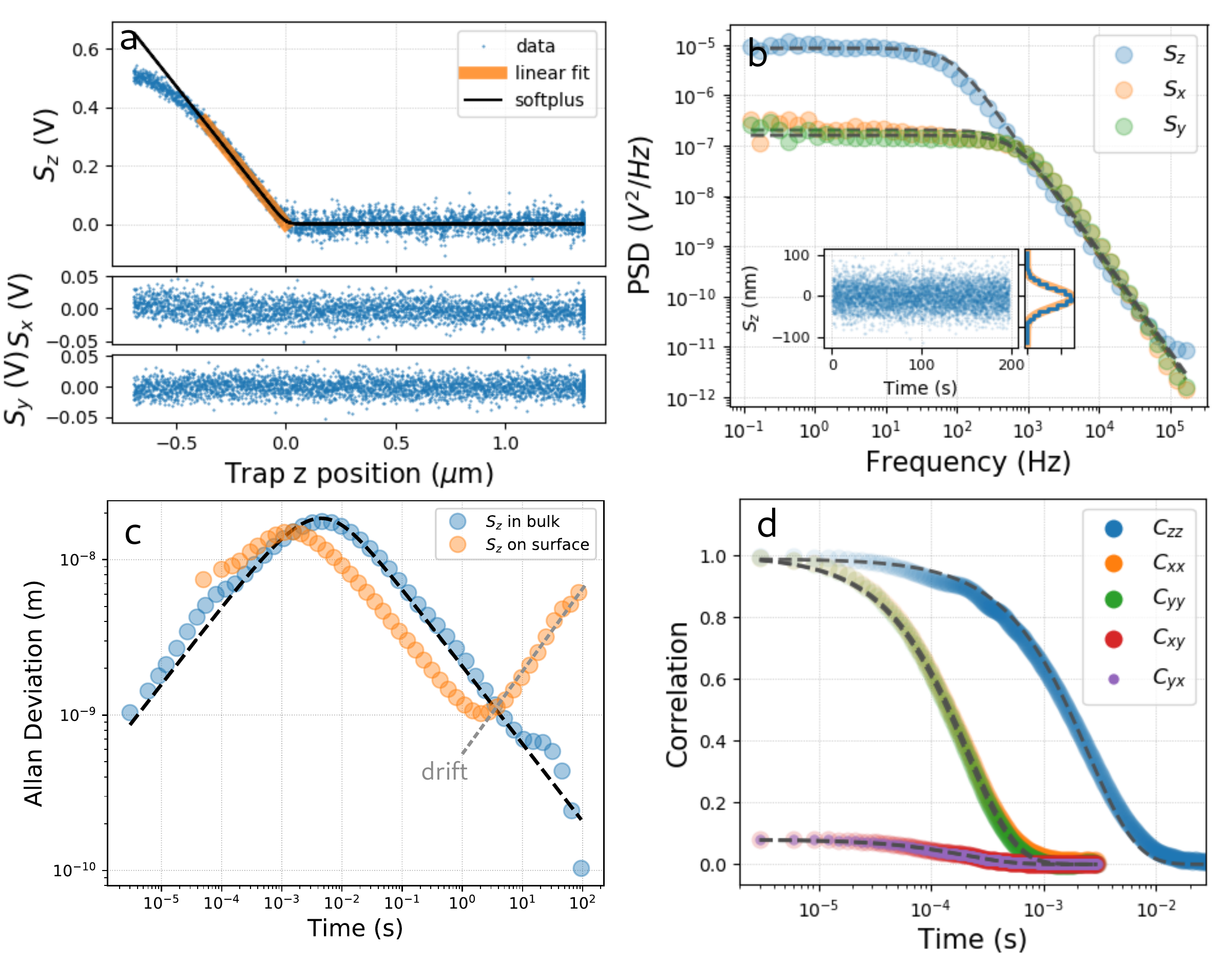}
\caption{Characterization of the trapped particle.
\textbf{a)} Indentation measurement. The signals $S_{z,x,y}$ are recorded while the optical trap is moved along $z$. The particle starts interacting with the glass surface at $z=0$. The $S_z$ signal is linear in the region highlighted in orange. 
\textbf{b)} Power spectral density (PSD) of the signals $S_{z,x,y}$ acquired from a particle trapped in the liquid bulk. The dashed lines correspond to Lorentzian fits to the data. The inset shows the signal $S_z$ in nm, together with its probability density (blue histogram), fit by a gaussian distribution (orange line).
\textbf{c)} Allan deviation for a particle trapped in bulk (blue points), and in contact to the surface (orange points). The black dashed line corresponds to the fit of the analytical expression \cite{van2018quantifying} for a particle in a harmonic potential. When the particle is in contact with the glass surface, drift becomes relevant for measurements integrating for more than 1 s.
\textbf{d)} Auto- and cross-correlation functions of the signal $S_{z,x,y}$ for a particle trapped in the liquid bulk. The dashed lines correspond to single-exponential fits.
}
\label{fig_2}
\end{figure}

Despite its complex geometry, the particle in the trap behaves not too differently from the standard spherical dielectric beads commonly used in OT. We characterize and quantify its behavior in fig.\ref{fig_2} by analyzing its overdamped Langevin dynamics (since inertia does not play a role at low Reynolds number). In fig.\ref{fig_2}a we perform a distance-displacement measurement: the signals $S_{z,x,y}$ are recorded while the trap position slowly scans the axial direction and contacts the rigid substrate surface. The resulting $z$-displacement of the tip within the trap can be approximated by a linear curve for displacements up to 400 nm. 
At the same time, the $S_x$ and $S_y$ signals remain constant, indicating that the particle $x,y$ position is only slightly perturbed when exploring the extreme axial region of the optical potential.
In order to calibrate the instrument, we extract the sensitivity $\beta$ (0.94 mV/nm) from the linear region of the $z$-displacement curve in fig.\ref{fig_2}a, by which one can map the signals from Volts to physical distance (see e.g. the inset of fig.\ref{fig_2}b). Fitting the power spectrum shown in fig.\ref{fig_2}b with a Lorentzian function \cite{berg2004power}, we obtain the trap stiffness $k_z$ (3 fN/nm) and the particle drag $\gamma_z$ (6$\,10^{-6}$ pN s/nm).
To test the calibration further, we compare the resulting drag coefficient $\gamma_z$ with the theoretical value $\gamma_\parallel$ obtainable for microscopic cylinders \cite{tirado1979translational} (see SI sec.\ref{fig_SI_theodrag}), which we find in the range $\gamma_\parallel = 6-9 \,10^{-6}$ pN s/nm, validating our experimental value. Also, if we consider the theoretical value of the drag coefficient $\gamma_\perp$ for a cylinder translating in the plane $x,y$ we can calibrate the transverse dimensions of the OT as well. This allows us to obtain images of the sample from the three displacement signals $S_z,S_x,S_y$.

\begin{figure}[h]
\centering
\includegraphics[width=0.7\textwidth]{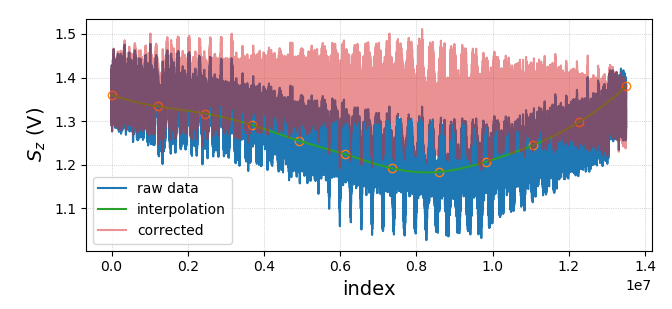}
\caption{Drift correction on the flatten version of the image. To remove the low frequency content due to the drift during the acquisition in the image, the flatten image (raw data, blue) is interpolated by cubic spline (points and green line). This interpolation is then sbtracted from the raw data (corrected, red).}
\label{fig_SI_rminterp}
\end{figure}

In fig.\ref{fig_2}c, we trace the Allan deviation \cite{lansdorp2012power,gibson2008measuring} of $S_z$, which quantifies the accuracy of the measured mean value of the signal as a function of the integration time used. When measured in bulk, the Allan deviation can be fit by an analytical expression for a spherical particle in a harmonic potential \cite{van2018quantifying}. The fit provides a trap stiffness $k_z$ (4 fN/nm) compatible with the previous results (while the drag $\gamma_z=11\,10^{-6}$ pN nm/s is far from the value obtained above, because of the hypothesis of spherical geometry). 
The Allan deviation peak indicates that the particle requires $\sim 10$ ms to explore the trap volume and that $\sim 4$ s of integration are required to reach 1 nm accuracy. 
When measured with the particle in contact with the surface, the Allan deviation indicates that drift in our setup starts to deteriorate the $z$-accuracy for integration times longer than 1 second. In the following, when imaging the sample surface with these particles, as a compromise between $z$-measurement accuracy and imaging time, we fix the time spent on each pixel to 80 ms, expecting an accuracy in $z$ of few nanometers (see SI fig.\ref{fig_SI_rminterp} for the implementation of drift correction).

\begin{figure}
\centering
\includegraphics[width=0.60\textwidth]{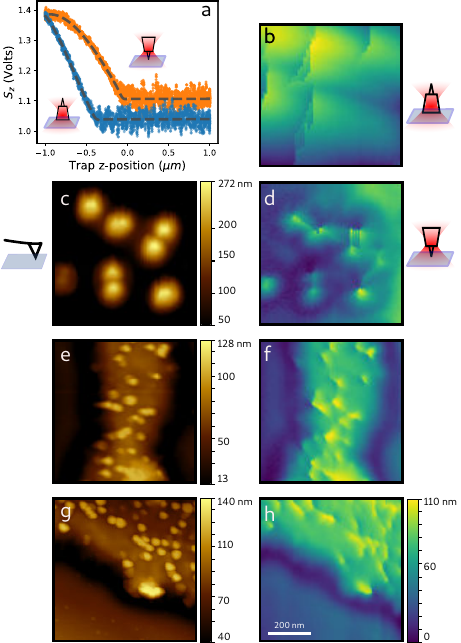}
\caption{AFM and PhFM imaging of rigid samples. 
\textbf{a)} $z$ position-displacement curves for the two possible particle orientations. When the tip faces the surface (orange points), the contact point appears before ($\sim 400$ nm) than when the tip is in the opposite direction (blue points). The shape and relative values (in Volts) of the two curves differ and allow systematic recognition of the tip orientation. The dashed lines correspond to a fit to a piece-wise polynomial function.
\textbf{b)} Image of 100 nm random features on glass obtained with the tip not facing the sample. \textbf{c)} AFM image and \textbf{d)} PhFM image (with tip facing the sample) of the same region. \textbf{e,d)} AFM and \textbf{f,h)} PhFM images of the same regions. Details of 80 nm can be resolved (see SI fig.\ref{fig_SI_resolution}). All the scans cover an area of 1x1 $\mu$m$^2$.
}
\label{fig_3}
\end{figure}

Additionally, the auto- and cross-correlations of the signals carry interesting information about the linear and angular fluctuations of the elongated particle \cite{volpe2006torque, marago2008femtonewton, irrera2011size}. 
After correcting for a small crosstalk between the signals $S_x$ and $S_y$ (see SI sec.\ref{SIsec_crosstalk}), we find that both the auto-correlations $C_{ii}(\tau) = \langle S_i(t)S_{ii}(t+\tau)\rangle$ ($i = x,y,z$) and the transverse cross correlation $C_{xy}(\tau)$ can be fit by a single exponential (fig.\ref{fig_2}c).
The values of the exponential rates of $C_{xx}$, $C_{yy}$, and $C_{xy}$ (corresponding to $\omega_i = k_i/\gamma_i$ for $C_{ii}$ ($i=x,y$), and to $\Omega_\Theta = k_\Theta/\gamma_\Theta$ for $C_{xy}$, where $\Theta$ indicates the tilt angle, see fig.\ref{fig_1}c) are found within 3\% relative error. 
Considering the theoretical value for the angular drag coefficient of a cylinder with the dimensions compatible with our particle ($\gamma_\Theta=8.8$ pN nm s, see SI sec.\ref{SIsec_theory_drag}), we can estimate from the exponential fit of $C_{xx}$ and $C_{yy}$ ($\Omega=4530\,s^{-1}$) the value of the angular stiffness ($k_\Theta=4\,10^{4}$ pN nm, see SI sec.\ref{SIsec_correlations}). The tilt angular fluctuations of the trapped particle can be estimated from $\sigma_\Theta = \sqrt{k_B T k_\Theta^{-1}} \sim 0.6$ degrees.
In conclusion, these results show that the particle remains stably trapped vertically in the laser focus, and that its axial position can be accurately measured by the signal $S_z$.

To validate our approach, we employ the nano fabricated tips to image structures on a hard substrate which can be reliably imaged by conventional AFM for comparison. To this end we etched the glass coverslip surface masked by randomly adhered 100 nm polystyrene beads. We find that the scattering due to the sample is negligible with respect to that from the trapped particle, due to its micrometer-scale size. Therefore we do not implement here the correction successfully proposed by Friedrich et al. for more optically thick samples \cite{friedrich2015surface}.
The results are shown in fig.\ref{fig_3}. First, for each particle trapped, we need to assess whether the tip faces the sample. We have found that this can be done easily and systematically by running a set of $z$-displacement measurements (fig.\ref{fig_3}a) where, between each measurement, the laser trap is briefly blocked to let the particle to reorient. The curves corresponding to the two configurations (tip up, tip down) are well recognizable from their shape and relative position. 

\begin{figure}[h]
\centering
\includegraphics[width=0.9\textwidth]{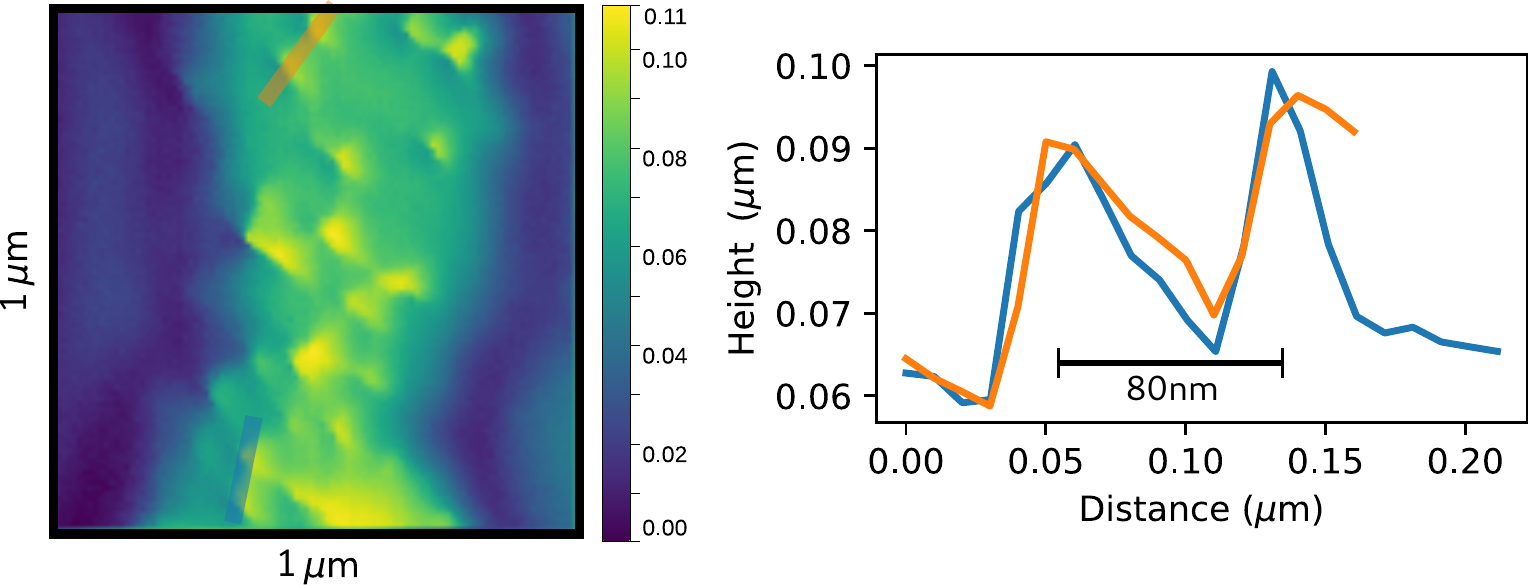}
\caption{Two height profiles (right) are extracted at the respective positions indicated by the colored bars in the image (left) of the glass etched structures (as in fig.\ref{fig_3}f).}
\label{fig_SI_resolution}
\end{figure}

We then record the signals $S_{z,x,y}$ while scanning the sample surface sequentially at discrete positions spaced by 10 nm, which correspond to the position of each pixel.
The images in fig.\ref{fig_3}b and fig.\ref{fig_3}d correspond to the image of the same region obtained with the two tip configurations: the details of the sample appear only when the tip faces the sample. The image of the same region obtained by AFM (fig.\ref{fig_3}c) confirms the one obtained by PhFM (fig.\ref{fig_3}d). Two more comparisons between images of the same sample obtained by AFM and PhFM are shown in fig.\ref{fig_3}e-f and fig.\ref{fig_3}g-h, and indicate that features of at least 80 nm (see SI fig.\ref{fig_SI_resolution}) can be well resolved with the nanofabricated tips by PhFM. Given the stiffness of the optical trap ($k_z = 3$ fN/nm), the maximum force applied by the tip on the sample remains lower than 0.4 pN. 
Interestingly, the image obtained from the transverse tip displacement ($S_x,S_y$) highlights the borders of the objects scanned (SI fig.\ref{fig_SI_xycorrection}).

\begin{figure}
\centering
\includegraphics[width=0.80\textwidth]{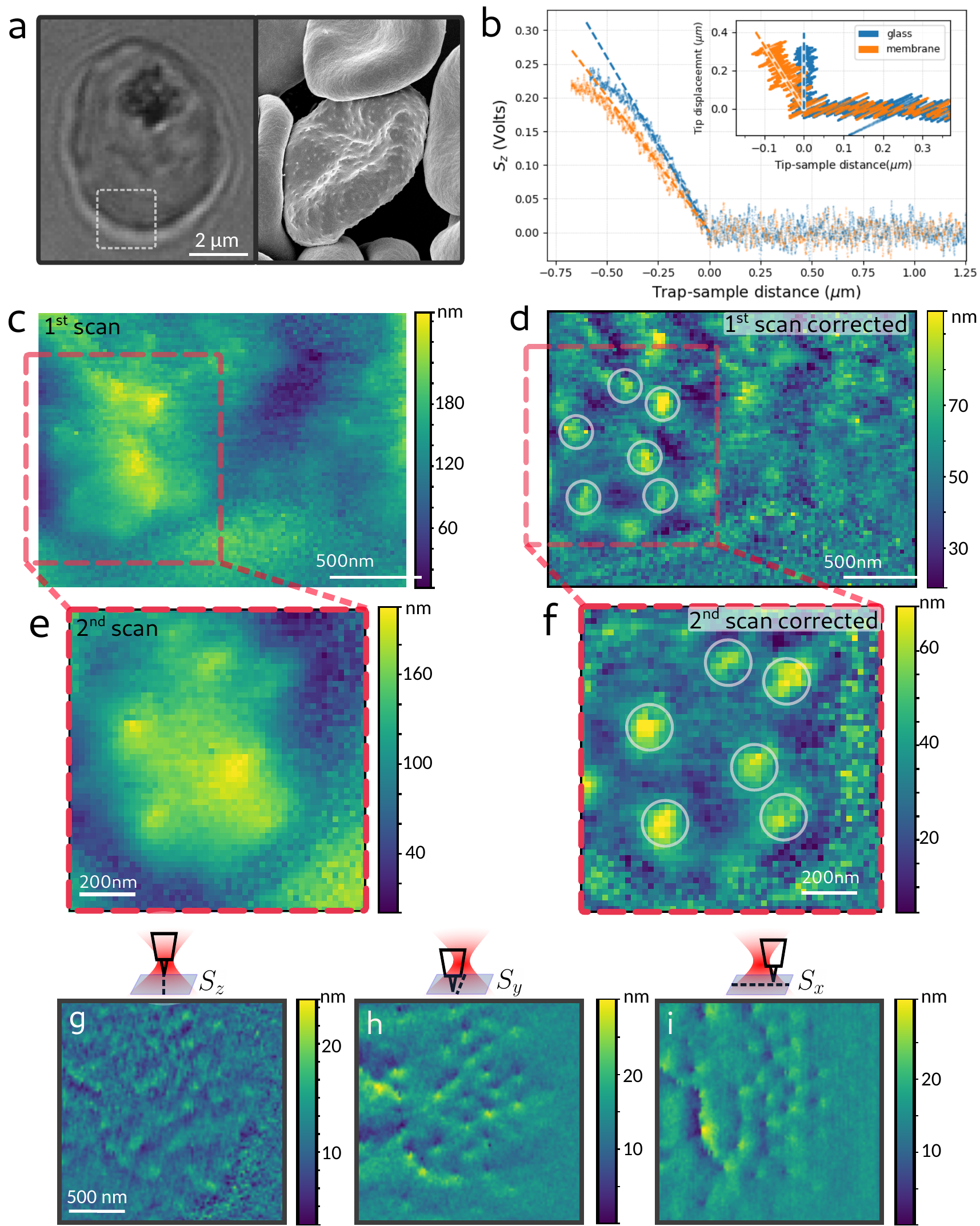}
\caption{PhFM of the membrane of living malaria infected red blood cells presenting knobs on its surface. 
\textbf{a)} Optical (left) and SEM (right \cite{knobsSEMwiki}) image of an infected red blood presenting knobs in its surface. A region scanned by PhFM is shown in the optical image.
\textbf{b)} Axial position-displacement curves obtained on the hard glass surface and on the cell membrane. The linear fits after contact are shown by dashed lines. Inset: corresponding indentation curves, where the curve corresponding to the glass surface is set vertical.
\textbf{c)} Image obtained from the axial displacement signal $S_z$ in a first scan of an infected cell, and  \textbf{d)} its high-pass filtered version. 
\textbf{e)} Image obtained in a subsequent scan in the region highlighted in red, and \textbf{f)} its filtered version. In d) and f) the same protrusions ascribable to knobs are marked by circles, showing that their position is the same in the two scans. 
\textbf{g-i)} Scan on a different infected cell, where the images are obtained from the signals $S_z,S_x,S_y$ as indicated, and high-pass filtered. Despite the limited axial resolution in $S_z$ (g), the presence of the knobs is retrieved in the transverse signals (h-i).}
\label{fig_4}
\end{figure}

To demonstrate the potential of the technique in scanning intact living cells (without artificial hardening the sample, as routinely done in AFM to produce a sufficient probe displacement) we now image the external membrane of intact red blood cells infected by the malaria parasite. 
%To further test our particles on soft biological material, we image the external membrane of live red blood cells infected by the malaria parasite.
When \textit{Plasmodium falciparum} infects a red blood cell, it exports newly expressed proteins (such as PfEMP1 and KHARP) to the host cell membrane, altering its otherwise smooth topography. As a result, small features of about 100 nm, termed \textit{knobs}, appear on the surface, stiffening the membrane and causing increased cell adherence and eventually vascular flow nuisances \cite{sanchez2019single} (fig.\ref{fig_4}a). 
The infected red blood cells are immobilized on a poly-L-lysine (PLL) coated glass cover slip, which also applies tension to the membrane. The cells are otherwise not treated biochemically. 

\begin{figure}[h]
\centering
\includegraphics[width=0.99\textwidth]{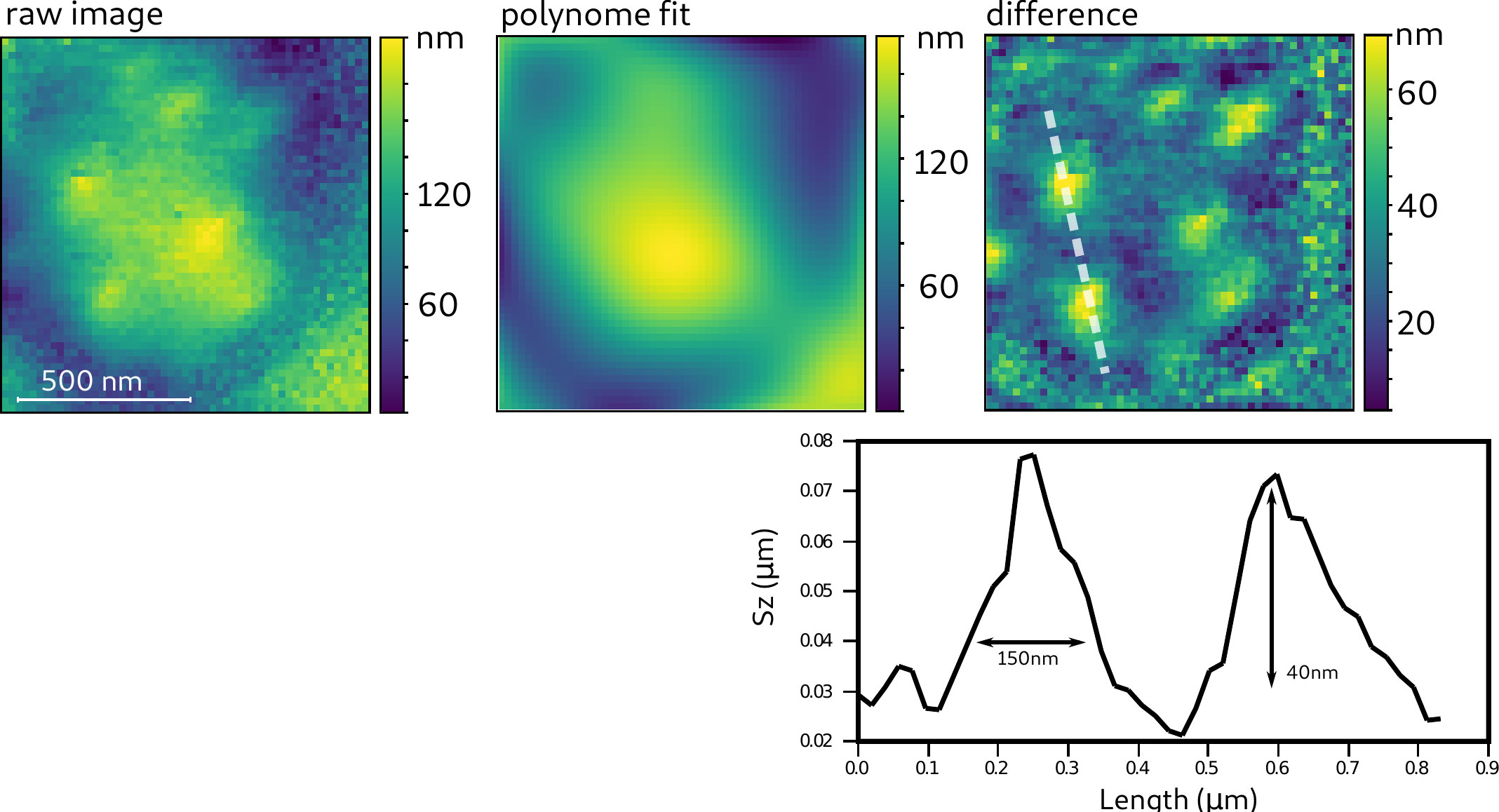}
\caption{Image filtering procedure. The raw image (same as fig.\ref{fig_4}e) is fit by a 2D polynomial function (degree 5 here) to obtain the low frequency content. The high-pass filtered image is found by the difference between the fit and the original image. The profile of the knobs along the dashed line is shown in the lower panel.}
\label{fig_SI_filtering}
\end{figure}

\begin{figure}[h]
\centering
\includegraphics[width=0.7\textwidth]{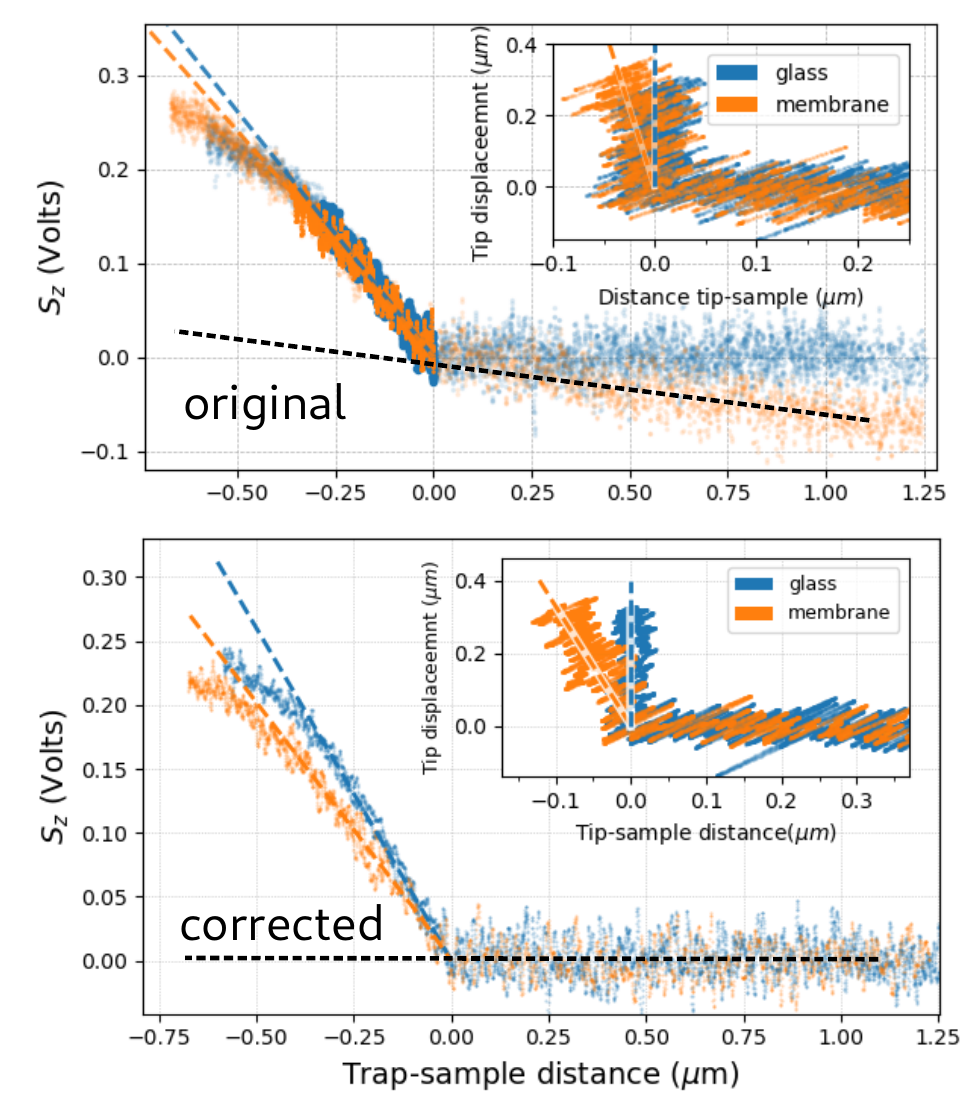}
\caption{Correction applied to the indentation measurement of the infected red blood cell. Top: due to scattering from the cell, the original $S_z$ curve obtained approaching the optical trap to the membrane (orange) displays systematically a tilt (dashed line) in the region where the trap is far from the membrane surface. Bottom: (same as fig.\ref{fig_4}b in the main text) the tilt is corrected by subtraction.}
\label{fig_SI_indent_correction}
\end{figure}

\begin{figure}[h]
\centering
\includegraphics[width=0.75\textwidth]{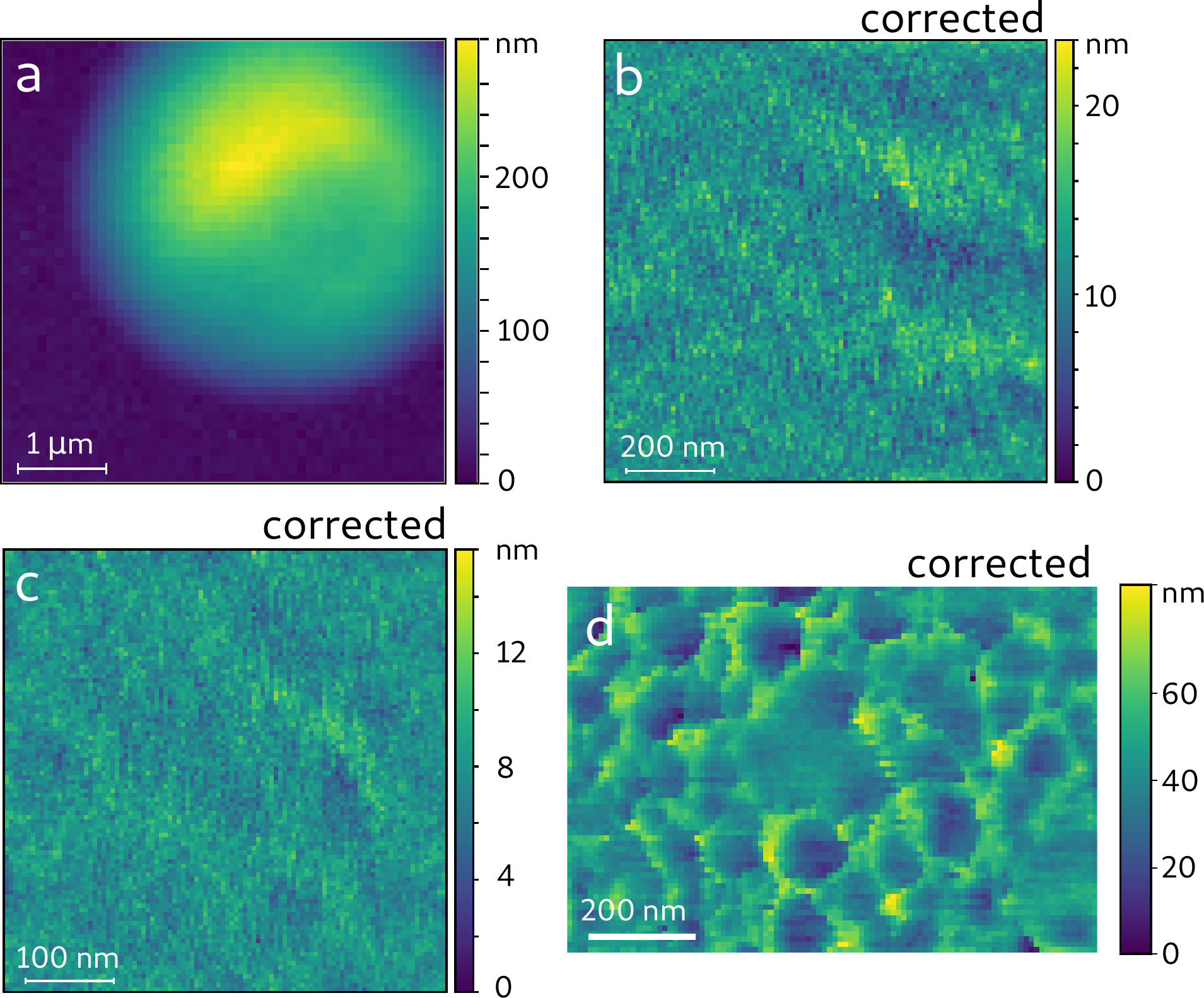}
\caption{Images of non-infected red blood cells. The cells are prepared as ghosts \cite{schwoch1973preparation} to exchange their optically thick hemoglobin content with external buffer. Their global structure is otherwise unchanged. They are immobilized by PLL coating on the glass coverslip, which builds tension on the membrane and deforms the normal bi-concave conformation of the cell into a dome-shape. Each image corresponds to a different cell. 
\textbf{a)} The 5x5 $\mu m^2$ region shows the top part of the cell. The image is formed from the uncorrected $S_z$ calibrated signal. 
\textbf{b)} Corrected (high pass filtered) scan of 1x1 $\mu m^2$, \textbf{c)} Corrected scan of 0.5x0.5 $\mu m^2$. 
\textbf{d)} A particular case of a scan of a red blood cell with structures that we could not easily see on other cells. We speculate that the valleys (darker regions) observed here could correspond to regions of the membrane between the cytoskeleton network, which holds the membrane on the cytosolic side, and which could become visible in a narrow region of the membrane tension applied by PLL. In any case, these structures are clearly different from the knobs of infected cells shown in fig.\ref{fig_4}.}
\label{fig_SI_RBCscans}
\end{figure}

In fig.\ref{fig_4}b we first show the results of membrane indentation measurements. The signal $S_z$ is shown as a function of the axial position of the optical trap, approaching both the cell membrane and the rigid glass surface. Taking as reference the curve obtained on glass, the inset shows that the membrane is indented by a maximum of 100 nm when the tip axial displacement is 300 nm.
The calibration provides here an axial stiffness $k_z = 1.5$ fN/nm, therefore the maximum force applied on the membrane by the tip is 0.45 pN. These measurements have been corrected by subtracting a background due to scattering, as shown in SI fig.\ref{fig_SI_indent_correction}.
We then scan a 2x2 $\mu m^2$ region of the cell far from the location of the optically dense hemozoin (a byproduct of hemoglobin digestion) and of the parasite (fig.\ref{fig_4}a).
When removing the low frequencies by 2D polynomial subtraction from the original image (fig.\ref{fig_4}c and SI fig.\ref{fig_SI_filtering}), small protrusions of $\sim 100$ nm diameter and $\sim 20$ nm height become clearly visible (fig.\ref{fig_4}d). 
A subsequent scan of the region highlighted in red in fig.\ref{fig_4}c-d confirms the positions and dimensions of these structures ($\sim150$ nm $\times 40$ nm,  fig.\ref{fig_4}e-f, SI fig.\ref{fig_SI_filtering}), which are compatible with knobs observed by electron microscopy and AFM on dead cells \cite{sanchez2019single, watermeyer2016spiral, rug2006role}. 
Images of similarly immobilized non-infected red blood cells show a smoother topography where such structures are absent, as expected (SI fig.\ref{fig_SI_RBCscans}). 
In fig.\ref{fig_4}g-i, we show the scan of a different infected live cell, where the axial resolution (fig.\ref{fig_4}g) is barely sufficient to resolve the knobs even after image filtering. However, information about the membrane topography can be retrieved from the transverse signals $S_x,S_y$ (fig.\ref{fig_4}h-i), which reflect the lateral shift of the particle in the trap due to the sample topography. This shows the rich information easily obtainable by PhFM (similar to lateral force microscopy in AFM \cite{mateAtomicscale_PRL1987}) the exploitation of which will be explored in future studies.

In conclusion, we have shown a simple nanofabrication protocol to massively produce optically trappable particles with sharp tips, and, characterizing their behavior in the trap, we have shown that they are well suited for scanning imaging. Employing the tips in PhFM, we obtain images of rigid substrates with resolution below 80 nm which compare well with those obtained by AFM. The potential of PhFM in scanning soft materials while applying sub-pN force is shown in imaging the 150 nm $\times$ 40 nm knobs on the membrane of red blood cells infected by \textit{P. falciparum}. Images are obtained from living cells, with a maximum force of a fraction of a pN, introducing no visible artefacts. Other scanning techniques mainly require chemical fixation to rigidify the sample (therefore killing the cells), and need to apply minimum forces of tens to hundreds of pN to obtain a sufficient probe deflection and a good signal to noise ratio in the image (a notable alternative is scanning ion conductance microscopy \cite{korchev1997scanning}).
In our implementation of PhFM, we find that a critical factor is the passivation of the probe (currently we use incubation in bovine serum albumine). Aspecific binding of the probe to the sample often terminates the scan, and the low force of the optical trap is not always sufficient to free the particle. Novel strategies have to be explored in this regard.
The speed of imaging depends on the time required for the probe to accurately sample each pixel (a 100x100 pixels image requires 13 minutes). To decrease this characteristic time (equal to $\gamma/k$), the trap stiffness $k$ can be increased via the laser power, but this is not always compatible with delicate samples. Alternatively, smaller particles could be engineered from higher index materials, in order to decrease the drag while keeping high stiffness. A good candidate is TiO$_2$ \cite{ha2019single}. Interestingly, using a birefringent material (like quartz or TiO$_2$) also allows for torque manipulation \cite{hategan2003adhesively, deufel2007nanofabricated, pedaci2011excitable}.
The photonic force microscope and optical tweezers are versatile and multi-functional instruments, which allow mechanical and dynamical measurements, together with manipulation at the micro- and nano-scale. Our results open the way to new PhFM  development for high-resolution imaging.

\section*{Acknowledgements}
This paper is dedicated to the memory of our colleague Jelle van der Does.
We are grateful to to Luca Costa, Pierre Emmanuel Milhiet, and Felix Rico for fruitful inputs and discussions, to Didier Portran and Viviana Claveria for their help in sample and cell preparations, to Frederic Pichot for the micro fabrication, Roland Teissier for the development of the LIL setup, and Michel Ramonda for the AFM images.
We acknowledge funding from the European Research Council under the European Union’s Seventh Framework Programme (FP/2007–2013)/ERC Grant Agreement No. 306475. 
The CBS is a member of the France-BioImaging (FBI) and the French Infrastructure for Integrated Structural Biology (FRISBI), 2 national infrastructures supported by the French National Research Agency (ANR-10-INBS-04-01 and ANR-10-INBS-05, respectively).

\section*{Authors contribution}
RD, BC developed the nanofabrication protocol.
ZS, EB, FP developed OT hardware and software.
RD, AN, MA, CBB prepared the cellular samples.
RD, ZS, EB, AN performed OT measurements.
FP, OM, ZS, EB, MA analyzed the PhFM data.
All authors wrote the manuscript.

\section*{Competing Interests statement}
The authors declare no competing interests.

\printbibliography

%%%%%%%%%%%%%%%%%%%%%%%%%%%%%%%%%%%%%%%%%%%%%%%%%%
\pagebreak

\section*{Methods}

\subsection*{Nanofabrication protocol} 
\begin{figure}[h]
\centering
\includegraphics[width=0.75\textwidth]{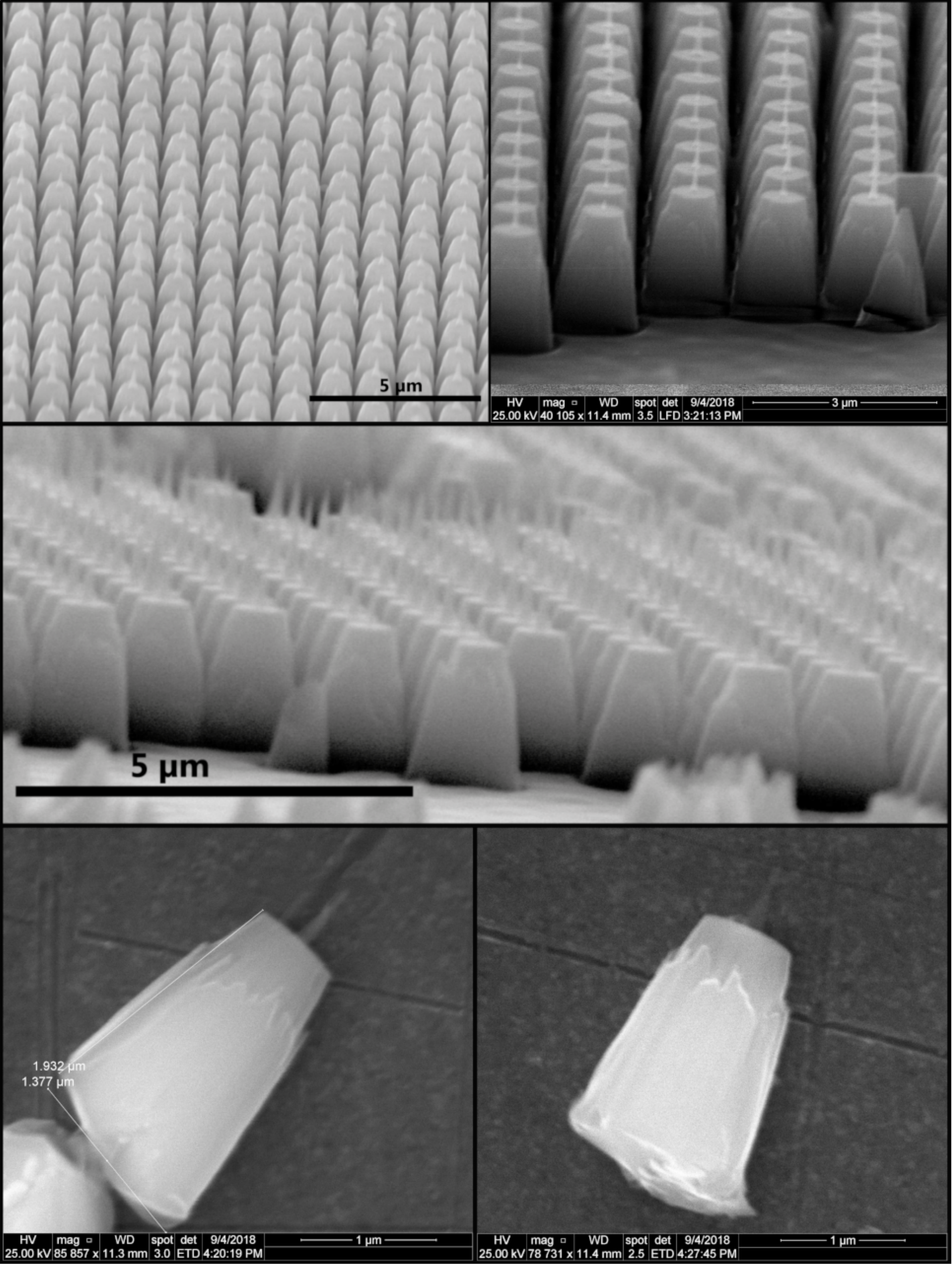}
\caption{SEM images of the particles at the end of the nanofabrication process, before and after cleavage.}
\label{fig_SI_SEMimages}
\end{figure}

The nanofabrication process is depicted in fig.\ref{fig_1}a and SI fig.\ref{fig_SI_SEMimages}.
One 4-inch x-cut single crystal quartz wafer is diced into 2x2 cm$^2$ chips using a dicing saw (DAD 3200, Disco, Japan). The substrate is cleaned with a Piranha solution (3:1 mixture of sulfuric acid (H$_2$SO$_4$) and hydrogen peroxide (H$_2$O$_2$)) and covered with 0.8 $\mu$m thick silicon dioxide deposited by Plasma Enhanced Chemical Vapour Deposition (PECVD) at 280$^o$C using a Corial D200 machine (Corial, France). 
Next, a 60 nm thick layer of chromium is deposited by electron gun evaporation (Univex 450, Leybold, Germany) on the whole surface. 
This layer acts as a hard mask for the subsequent plasma etching steps. On these materials stack, we spin-coated a thin layer of AZ 701 MIR photoresist (Merck Performance Materials GmbH, Germany) at 4000 rpm for 30 s, followed by soft baking at 95$^o$C for 1 min. The photoresist is diluted (2:1) to achieve a final thickness of 600 nm.

The quartz sample is then exposed by Laser Interference Lithography, as detailed in \cite{santybayeva2016fabrication}. Briefly, this mask-less technique uses subsequent laser exposures with interference fringes to produce a 2D array of dots on the sample (fig.\ref{fig_1}b). A divergent 405 nm laser beam strikes two adjacent mirrors facing each other at an angle $\theta$, creating two beams interfering at the plane of the photoresist-coated sample. 
A first exposure (of duration 130 s) imprints on the photoresist a set of parallel lines whose spatial period is 1.8 $\mu$m. 
In order to obtain an array of discs, we proceed with a second exposure with the substrate rotated by 90$^o$. After the exposures, the sample is baked at 110$^o$C for 1 min and developed with AZ 726 developer for 17 s under smooth agitation. The result is an array of discs with a spatial period of 1.8 $\mu$m and diameter of 1 $\mu$m. 

The next step consists in transferring the photoresist pattern onto the quartz substrate. To achieve the high anisotropy required by the cylindrical geometry, we use inductive coupled plasma reactive ion etching (ICP-RIE). The chromium layer is etched with Argon plasma in the absence of oxygen, using a Corial 250D ICP RIE (200 W RF and 400 W LF). The absence of oxygen prevents the photoresist from shrinking during etching, allowing the unprotected Chromium hard mask to be completely removed in 6 min with sufficient selectivity. Etching of the SiO$_2$ and quartz stack is achieved by a second ICP-RIE process using a mixture of CHF$_3$ (100 sccm) and O$_2$ (10 sccm), an internal pressure of 10 mTorr, and an incoming power of 120 W RF and 400 W LF. In these conditions, the etching rate of the quartz is about 100 nm/min. The cylinders obtained are 2.6 $\mu$m thick and slightly conical, with a wall angle of about 80 deg (fig.\ref{fig_1}b, SI fig.\ref{fig_SI_SEMimages}). 

In order to create the sharp tips, we use a thinning step by wet etching in hydrofluoric acid (HF), taking advantage of the large difference in etching rate between quartz and SiO$_2$ (1 and 200 nm/min, respectively \cite{williams2003etch}). The quartz substrate is entirely dipped in 5\% HF during 3 min, and due to the isotropic character of the SiO$_2$ etching, the profile of the SiO$_2$ cylinder forms a sharp tip. The remaining chromium discs are then removed by wet etching (in a mixture of perchloric acid (HClO4), and ceric ammonium nitrate (NH4)$_2$[Ce(NO$_3$)$_6$]). The last step consists of mechanically cleaving the particles using a blade (fig.\ref{fig_1}b). The particles are then collected and stored in aqueous solution.

\subsection*{Optical setup}
The optical tweezers setup, based on the one described in \cite{santybayeva2017optical}, is sketched in fig.\ref{fig_1}d. We use a linearly polarized infrared laser (1064 nm, 3W Azur Light Systems), whose intensity is adjusted and actively clamped to the suited value (20-60 mW at the objective input) by the combination of one half waveplate, one polarizer, and an acusto-optical modulator driven by a PID controller. The collimated beam is expanded to slightly overfill the back aperture of the microscope objective (Zeiss C-Apochromat 63x/1.2NA, water immersion), via a 1:1 telescope. A water immersion objective was chosen to minimize the z-dependent aberrations of the trap \cite{vermeulen2006optical}. The laser trap is generated in the liquid medium enclosed in a flow cell, prepared by two glass cover slips separated by one layer of parafilm and sealed. The samples used are positioned on the bottom glass surface. Due to the cylindrical geometry, the particle remains vertical in the trap, with its geometrical axis parallel to the laser propagation direction. The laser trap is linearly polarized, which produces a torque on the birefringent particle \cite{la2004optical, pedaci2011excitable} that prevents rotation around the geometrical axis. 
The flowcell position is controlled in three dimensions by a piezo stage with nm resolution (P-517.3CD). 
The laser trap output is collected by a condenser, identical to the objective, and analyzed by forward-scattering interferometry. To detect the lateral ($x,y$) displacement of the trapped particle, we use a position sensitive detector (PSD, DL100-7-PCBA3 First sensor Inc.), located on a conjugate plane of the back focal plane of the condenser. In parallel, the axial particle displacement ($z$) is measured by a photodetector (DET 10N/M, Thorlabs). This allows independent control of the numerical aperture at the two detector planes, maximizing both of their sensitivities (the aperture is maximal for the PSD, while it is reduced by an iris placed on-axis before the photodetector) \cite{dreyer2004improved, rohrbach2003three}. 
The voltage signals of the optical detectors and of the piezo stage are acquired synchronously by a FPGA card at 200-500 kHz (PXI-7852R, National Instruments). The hardware-software interface is developed in Labview, and the following data analysis is performed in Python.

\subsection*{Sample preparation}

\subsubsection*{Etched glass structures}
The sample used in fig.\ref{fig_3} consists of structures directly etched onto a 24x60 mm$^2$ microscope coverslip. Polystyrene micro-beads (100 nm diameter) are deposited on the glass surface by spincoating (2000 rpm for 30 s). A plasma etching reactor (Corial 250D ICP RIE) is used to etch the coverslip, the beads playing the role of a mask. 
During the scan, passivation of the surfaces is achieved by 10 minutes incubation of bovine serum albumin (BSA) in the flow cell.

\subsubsection*{Red blood cells}
P. falciparum 3D7 strain, obtained from the Malaria Research and Reference Reagent Resource Center (MR4-BEI resources, MRA-102), was cultured in human erythrocytes obtained as donations from anonymized individuals from the French Blood Bank (Etablissement Français du Sang, Pyrénées Méditerranée, France; approval number 21PLER2016-0103) at 5\% hematocrit in RPMI 1640 medium (Gibco), supplemented with gentamycin at 20 $\mu$g/ml and 10\% human serum \cite{trager1976human}. The cultures were kept at 37$^o$C under a controlled trigaz atmosphere (5\% CO$_2$ , 5\% O$_2$ and 90\% N$_2$). To minimize knobless parasite mutants under in vitro culture, mature parasites were isolated using gelatin floatation on a weekly basis \cite{pasvol1978separation}. The synchronization of the parasites was achieved by purification of mature parasites on a Variomacs column \cite{mata2014magnetic} followed by 5\% sorbitol \cite{lambros1979synchronization} treatment 2hours post-invasion. Mature trophozoites were collected by passage through a VarioMACS column (Miltenyi Biotec; $>95\%$ infected erythrocytes) and resuspended in complete culture medium under 5\% O$_2$ controlled atmosphere.

%%%%%%%%%%%%%%%%%%%%%%%%%%%%%%%%%%%%%%%%%%
\clearpage

\vspace{50pt}

\begin{center}
\textbf{\Large \textbf{Supplementary Information}\\\large High resolution photonic force microscopy based on sharp nano-fabricated tips}

\vspace{50pt}

Rudy Desgarceaux*, Zhanna Santybayeva*, Eliana Battistella, Ashley L. Nord, Catherine Braun-Breton, Manouk Abkarian, Onofrio M. Maragò, Benoit Charlot$^{\dagger}$, Francesco Pedaci$^{\dagger}$

\end{center}

\tableofcontents
\pagebreak

\section{Cross talk correction}
\label{SIsec_crosstalk}

\begin{figure}[h]
\centering
\includegraphics[width=0.5\textwidth]{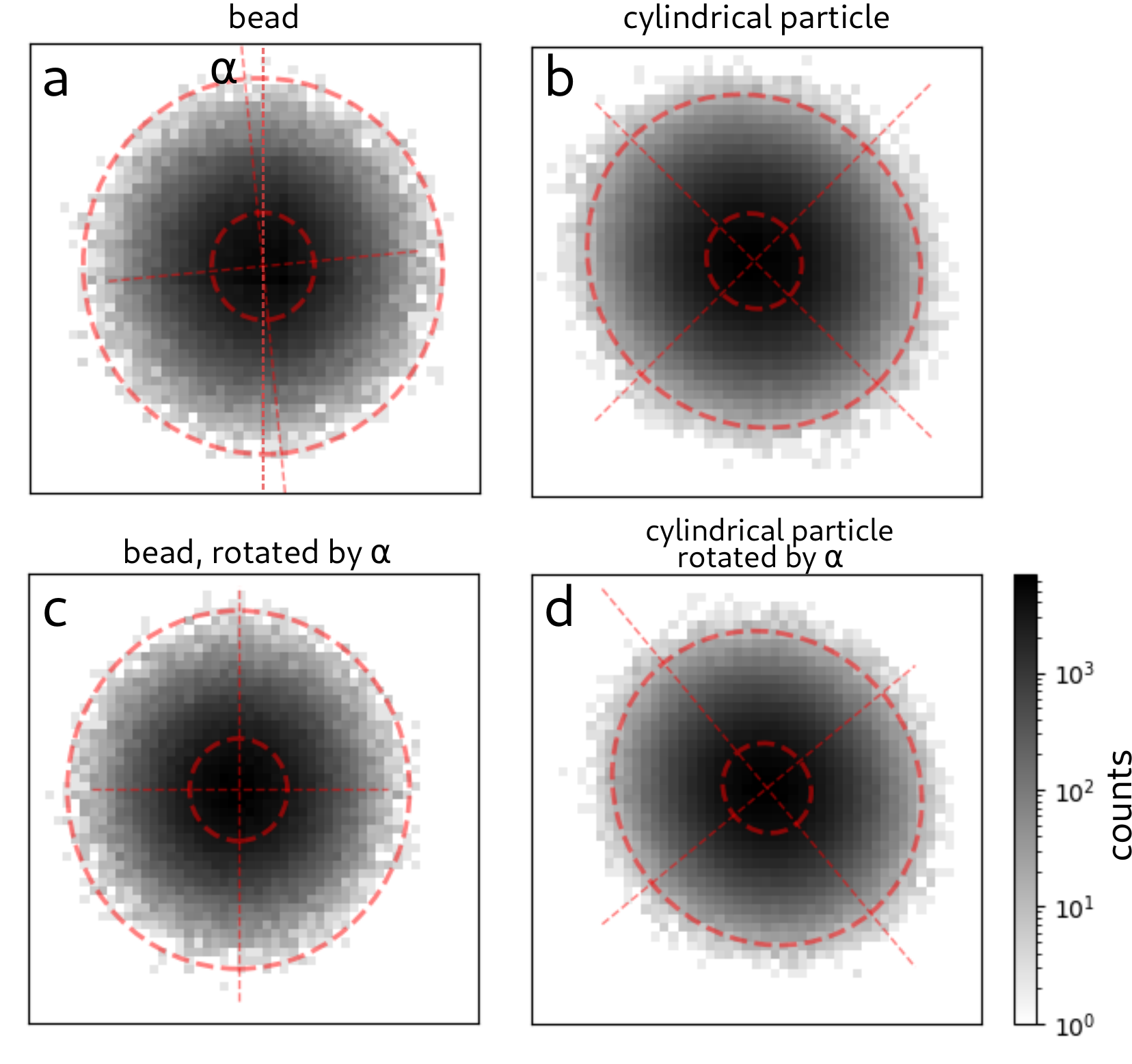}
\caption{Distributions of particle positions in the $x,y$ plane and Gaussian fit. 
\textbf{a)} Distribution of points visited in the plane $x,y$ by a 1 $\mu$m diameter bead. A two dimensional Gaussian fit finds the widths and angle (here $\alpha_{\mbox{\footnotesize{bead}}}=5.6^o$) of the distribution. 
\textbf{b)} Same for a nanofabricated particle with tip. 
\textbf{c) d)} Corresponding corrections of the cross talk by rotation of $-\alpha_{\mbox{\footnotesize{bead}}}$.
}
\label{fig_SI_crosstalk}
\end{figure}

To correct for the residual cross talk between the signals $S_x$ and $S_y$, we proceed as follows. We first quantify the cross talk by trapping a 1 $\mu$m diameter bead in bulk. We fit the corresponding points in the $S_x, S_y$ plane by a Gaussian model (based on the expectation-maximization algorithm implemented in python sklearn\footnote{https://scikit-learn.org/stable/modules/mixture.html}, see fig.\ref{fig_SI_crosstalk}a). The fit provides the two standard deviations ($\sigma_1, \sigma_2$) along the two main axis of the elliptical pattern as well as its angle $\alpha_{\mbox{\footnotesize{bead}}}$. 
Our tests show that a ratio $\sigma_1/\sigma_2=1.001$ is detectable by the algorithm, using an order of $10^6$ points as in the experiment.
For the bead, we find $\sigma_1/\sigma_2 = 1.047$. It is then possible to correct for the cross talk by rotating the points by an angle $-\alpha_{bead}$ (fig.\ref{fig_SI_crosstalk}c). Trapping a nanofabricated particle, the fit to the probability distribution provides $\sigma_1/\sigma_2 = 1.087$ (fig.\ref{fig_SI_crosstalk}b), and an angle $\alpha_{\mbox{\footnotesize{cyl}}}$ larger than $\alpha_{\mbox{\footnotesize{bead}}}$. The final distribution of points for the nanofabricated particle, corrected by the angle $\alpha_{\mbox{\footnotesize{bead}}}$ is shown in SI fig.\ref{fig_SI_crosstalk}d.

%\pagebreak

\section{Correlation analysis}
\label{SIsec_correlations}
Following the works that analysed the behavior of elongated and cylindrical nano-particles in optical traps at low Reynolds number \cite{volpe2006torque, marago2008femtonewton, irrera2011size}, we can summarize the following results.
The Langevin equations for the linear displacement of the center of mass ($X,Y,Z$) and for the angular tilt ($\Theta_x, \Theta_y$) can be written as 
\begin{eqnarray}
\label{eq_diffeqs}
\partial_t X(t) & = & - \omega_x X(t) + \sqrt{2k_BT\gamma_x}\, \xi_x(t) \nonumber \\
\partial_t Y(t) & = & - \omega_y Y(t) + \sqrt{2k_BT\gamma_y}\, \xi_y(t) \nonumber \\
\partial_t Z(t) & = & - \omega_z Z(t) + \sqrt{2k_BT\gamma_z}\, \xi_z(t) \\
\partial_t \Theta_x(t) & = & - \Omega_x \Theta_x(t) + \sqrt{2k_BT\gamma_{\Theta_x}}\, \xi_{\Theta_x}(t) \nonumber \\
\partial_t \Theta_y(t) & = & - \Omega_y \Theta_y(t) + \sqrt{2k_BT\gamma_{\Theta_y}}\, \xi_{\Theta_y}(t) \nonumber
\end{eqnarray}
where the rates are defined by 
\begin{eqnarray}
\label{eq_rates}
\omega_x = k_x/\gamma_\perp & \Omega_x = k_{\Theta_x}/\gamma_{\Theta_x} \nonumber \\ 
\omega_y = k_y/\gamma_\perp & \Omega_y = k_{\Theta_y}/\gamma_{\Theta_y}  \\
\omega_z = k_z/\gamma_\parallel & \nonumber
\end{eqnarray}
and where $k_i$ is the stiffness ($i=x,y,z,\Theta_x,\Theta_y$), $\gamma_\perp, \gamma_\parallel$ are the linear drag coefficients (perpendicular and parallel to the geometrical axis of the particle), $\gamma_{\Theta_x}, \gamma_{\Theta_y}$ are the angular drag coefficients, $\xi_i(t)$ are uncorrelated noise sources, and $k_BT$ is the thermal energy.
The (auto and cross) correlation functions for these degrees of freedom can be written \cite{volpe2006torque} in general as 
\begin{equation}
    C_{ab}(\tau) = \langle a(t)\,b(t+\tau)\rangle
    \label{eq_xcorr_ab}
\end{equation} 
where $a, b=(X,Y,Z,\Theta_x,\Theta_y)$. The auto-correlation functions follow the differential equation
$\partial_\tau C_{aa}(\tau) = -(k_a/\gamma_a)\, C_{aa}\,$, with exponentially decaying solution 
\begin{equation}
\label{eq_xcorrtheory}
    C_{aa}(\tau) = \langle a(t)\,a(t+\tau)\rangle = \frac{k_BT}{k_a}\,e^{-(k_a/\gamma_a)|\tau|} \qquad\qquad a=(X,Y,Z,\Theta_x,\Theta_y).
\end{equation}
where the equipartition theorem implies that
\begin{equation}
C_{aa}(\tau=0) = \langle a^2(t) \rangle = var(a) = \frac{k_BT}{k_a} \qquad\qquad\qquad (a=X,Y,Z,\Theta_x,\Theta_y).
\end{equation}
(note the different units: $[C_{xx,yy,zz}]=m^2$, $[C_{\Theta_{x,y}}]=1$, $[k_{x,y,z}]=N/m$, $[k_{\Theta_{x,y}}]=Nm$).

In the small tilt angle regime, the calibrated signals $S_z, S_x, S_y$ (already mapped to meters) acquired from the detectors couple together the linear $X,Y,Z$ and the angular $\Theta_x, \Theta_y$ variables. If $L$ is the length of the particle, it can be written
\begin{eqnarray}
    S_x &=& X+L\Theta_x \nonumber \\
    S_y &=& Y+L\Theta_y \\
    S_z &=& Z \nonumber
\end{eqnarray}
The correlation functions of the calibrated acquired \textit{signals} can therefore be written as
\begin{eqnarray}
\label{eq_correlexperim}
C_{xx}(\tau) &=& \langle X(t)\,X(t+\tau)\rangle + L^2\langle \Theta_x(t)\,\Theta_x(t+\tau)\rangle \nonumber\\
C_{yy}(\tau) &=& \langle Y(t)\,Y(t+\tau)\rangle + L^2\langle \Theta_y(t)\,\Theta_y(t+\tau)\rangle \nonumber\\
C_{zz}(\tau) &=& \langle Z(t)\,Z(t+\tau)\rangle \\
C_{xy}(\tau) &=& L^2 \langle \Theta_x(t)\,\Theta_y(t+\tau)\rangle \nonumber \\
C_{yx}(\tau) &=& L^2 \langle \Theta_y(t)\,\Theta_x(t+\tau)\rangle
\nonumber
\end{eqnarray}
(note the difference in notation between the correlation of the linear physical displacement $C_{XX}$ of eq.\ref{eq_xcorr_ab}, and the correlation of the corresponding measured signal $C_{xx}$).
Combining eq.\ref{eq_diffeqs},\ref{eq_rates},\ref{eq_xcorrtheory},\ref{eq_correlexperim}, one can explicitly write the expressions of the correlation functions of the \textit{signals} as
\begin{eqnarray}
\label{eq_correlexp2}
C_{xx}(\tau) &=& \frac{k_BT}{k_x} \, \exp{[-\frac{k_x}{\gamma_\perp} \tau]} + L^2\frac{k_BT}{k_{\Theta_x}} \, \exp{[-\frac{k_{\Theta_x}}{\gamma_{\Theta_x}} \tau]} \nonumber \\
C_{yy}(\tau) &=& \frac{k_BT}{k_y}\,\exp{[-\frac{k_y}{\gamma_\perp} \tau]} + L^2\frac{k_BT}{k_{\Theta_y}}\,\exp{[-\frac{k_{\Theta_y}}{\gamma_{\Theta_y}} \tau]} \nonumber \\
C_{zz}(\tau) &=& \frac{k_BT}{k_z}\,\exp{[-\frac{k_z}{\gamma_\parallel} \tau]} \\
C_{xy}(\tau) &=& C_{xy}(0) \, \exp{[-\frac{k_{\Theta_y}}{\gamma_{\Theta_y}} \tau]} \nonumber \\
C_{yx}(\tau) &=& C_{yx}(0) \, \exp{[-\frac{k_{\Theta_x}}{\gamma_{\Theta_x}} \tau]} \nonumber
\end{eqnarray}
where $\tau\ge 0$, $C_{xy}(\tau) = C_{yx}(-\tau)$. For $\tau=0$, now one obtains for $x,y$ a combination of linear and angular displacement:
\begin{eqnarray}
    C_{xx}(0) = \langle S_x(t)^2 \rangle = var(S_x) = \frac{k_BT}{k_x}+L^2\frac{k_BT}{k_{\Theta_x}} \\
    C_{yy}(0) = \langle S_y(t)^2 \rangle = var(S_y) = \frac{k_BT}{k_y}+L^2\frac{k_BT}{k_{\Theta_y}} \nonumber
\end{eqnarray}

We note the following: 
1) experimentally (see fig.\ref{fig_2}d), we find that all the above correlation functions can be fit by a single exponential. The single-exponential rate of $C_{xx}$, $C_{yy}$, and $C_{xy}$ are equal (within 3\%). 2) from the exponential fit of $C_{xy}$ and $C_{yx}$, we find $\Omega_x = \Omega_y = \Omega = 5030 \, \frac{1}{\mbox{s}}$ (within 0.6\%). We can set $\gamma_{\Theta} = \gamma_{\Theta_x} = \gamma_{\Theta_y}$, so it results $k_{\Theta} = k_{\Theta_x} = k_{\Theta_y}$. Because $C_{xx}$ and $C_{yy}$ are very similar, we can set $C_\perp=C_{xx}=C_{yy}$, and $\omega_x = \omega_y$, so $k_\perp=k_x=k_y$. We call $C_\parallel=C_{zz}$, and  $k_z=k_\parallel$.
Therefore we can write a simplified version of eq.\ref{eq_correlexp2} as
\begin{eqnarray}
C_\perp(\tau) &=& \frac{k_BT}{k_\perp}\,\exp{[-\frac{k_\perp}{\gamma_\perp}\tau]} + \frac{k_BT}{k_\Theta}L^2\,\exp{[-\frac{k_\Theta}{\gamma_\Theta}\tau]} \nonumber \\
C_\parallel (\tau) &=& \frac{k_BT}{k_\parallel}\,\exp{[-\frac{k_\parallel}{\gamma_\parallel}\tau]} \\
C_{xy}(\tau) &=& C_{xy}(0)\exp{[-\frac{k_\Theta}{\gamma_\Theta}\tau]} \nonumber 
\end{eqnarray}
We now calculate the theoretical values of the drag coefficients for a cylinder with length $L$ and diameter $d$ similar to our particle ($L=1.2-1.7\,\mu$m, $d=0.7-1.2\,\mu$m, see SI sec.\ref{SIsec_theory_drag}):
\begin{eqnarray}
\label{eq_SI_dragvalues}
\gamma_\parallel &=& (8.8 \pm 1.9)\, 10^{-6} \,\, \mbox{pN s/nm}  \nonumber \\
\gamma_\perp &=& (1.4 \pm 0.3)\,10^{-5} \,\, \mbox{pN s/nm}  \nonumber \\
\gamma_{\Theta} &=& 8.81 \,\,  \mbox{pN nm s} 
\end{eqnarray}
From the value of $\Omega=\frac{k_\Theta}{\gamma_\Theta} = 4530\, s^{-1}$, we can estimate $k_\Theta = 4\,10^4$ pN nm. The variance of the tilt angular fluctuations is given by $\sigma^2_\Theta=\frac{k_BT}{k_\Theta}L^2$, corresponding to an angle $\sigma_\Theta/L = \sqrt{\frac{k_BT}{k_\Theta}} \sim 0.01$ rad (0.6 degrees).

%\newpage

\section{Theoretical drag of microscopic cylinders}
\label{SIsec_theory_drag}

\begin{figure}[ht]
\centering
\includegraphics[width=0.98\textwidth]{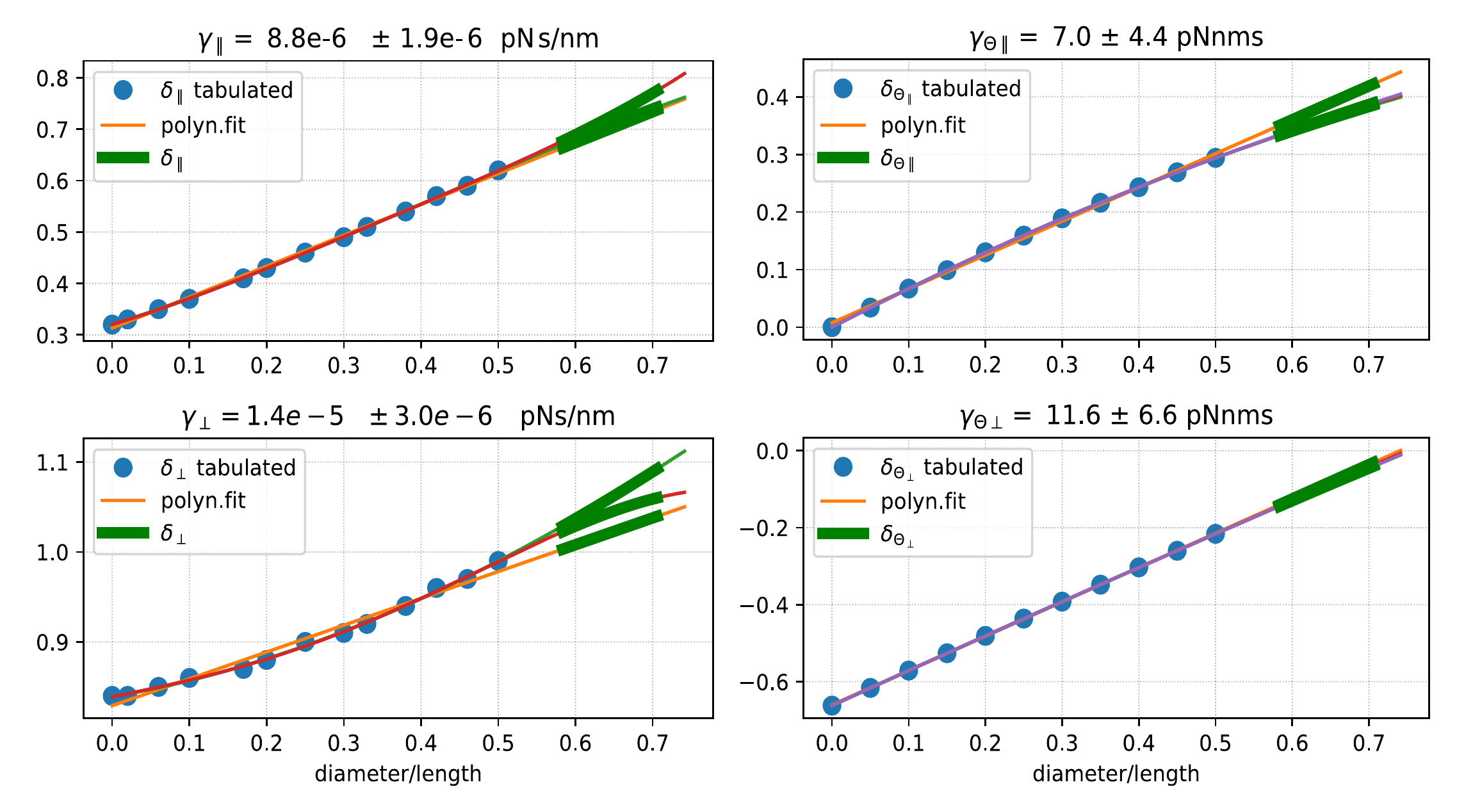}
\caption{Values of the parameters $\delta_i$ ($i=\perp,\parallel,\Theta_\perp,\Theta_\parallel$) used to calculate the theoretical drag coefficients for cylinders of different diameter and length. The particles used in this work lay in the region marked in green. The tabulated values \cite{tirado1979translational} are shown as blue dots, while their polynomial fits (with order 1 to 4) with lines. In each panel the title indicates the corresponding value of the drag $\gamma_i$.}
\label{fig_SI_theodrag}
\end{figure}
We use the results of \cite{tirado1979translational} to calculate the theoretical values of the drag coefficients of a cylinder of length $L$ and diameter $d$
\begin{eqnarray}
\gamma_\perp &=& \frac{4\pi\eta L}{\ln p + \delta_\perp} \nonumber \\
\gamma_\parallel &=& \frac{2 \pi \eta L}{\ln p + \delta_\parallel} \\
\gamma_{\Theta_\perp} &=& \frac{\pi \eta L^3}{3\,(\ln p + \delta_{\Theta_\perp}) } \nonumber \\
\gamma_{\Theta_\parallel} &=& 3.841\,\pi\eta L (d/2)^2 (1+\delta_{\Theta_\parallel}) \nonumber
\end{eqnarray}
where $\gamma_\perp$ and $\gamma_\parallel$ are the transverse and axial linear translational drag coefficients respectively, 
$\gamma_{\Theta_\perp}$ and $\gamma_{\Theta_\parallel}$ are the rotational drag coefficients for a rotation around an axis perpendicular and parallel to the symmetry axis respectively (note that we do not employ here $\gamma_{\Theta_\parallel}$, which is given just for completeness),
$\eta$ is the water dynamic viscosity, and $p=L/d$. The correction factors $\delta_\perp, \delta_\parallel, \delta_{\Theta_\perp}, \delta_{\Theta_\parallel}$ are tabulated in \cite{tirado1979translational} and plotted in SI fig.\ref{fig_SI_theodrag}, where their values are fit by polynomial expressions of different order (1 to 4). The region corresponding to a cylinder of dimensions compatible with our particles requires extrapolation the theoretical values by means of the fit, and is marked in green in the plots. The average and standard deviation of the corresponding drag coefficients $\gamma_i$ are indicated in each plot and used in eq.\ref{eq_SI_dragvalues}.

%\newpage
\section{Calibration and image reconstruction}

\label{SIsec_calibration_and_image}
\begin{figure}[ht]
\centering
\includegraphics[width=0.99\textwidth]{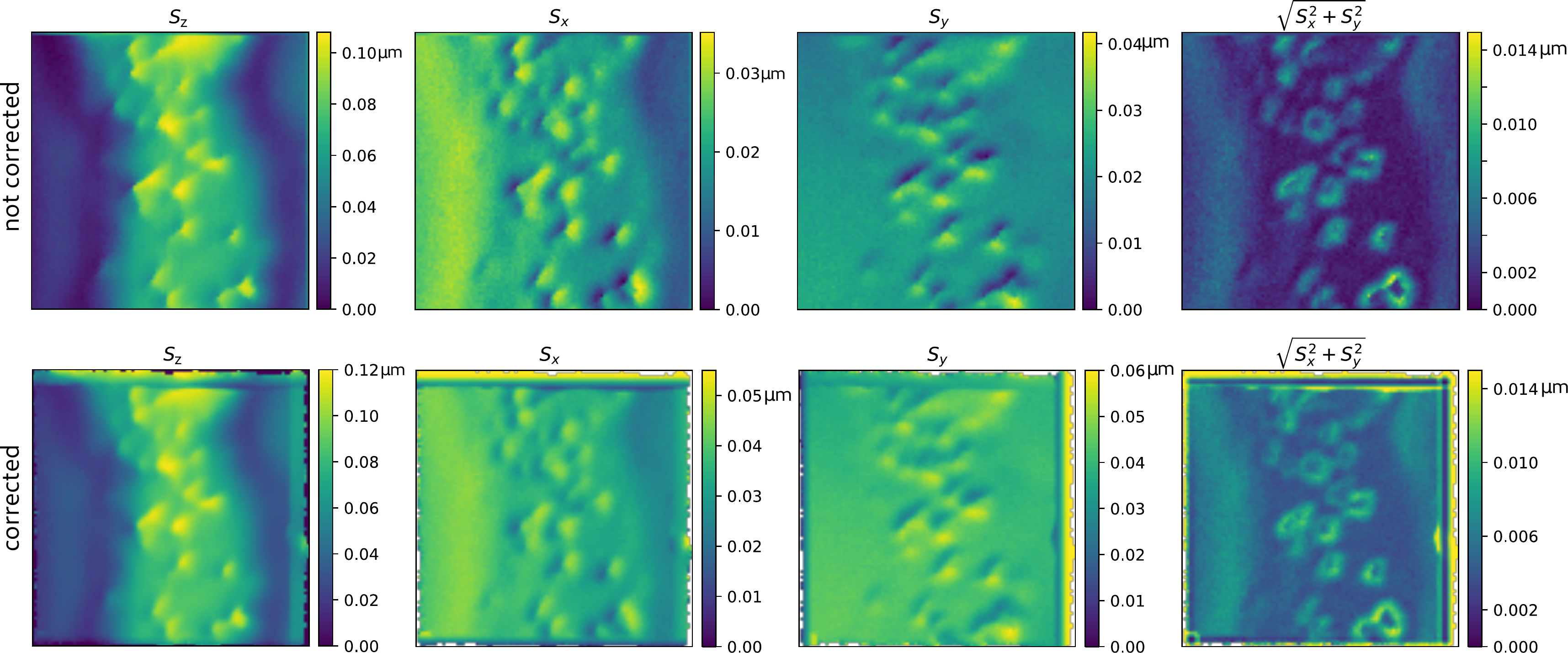}
\caption{Images from the acquired signals. Each column corresponds to the images obtained from the calibrated signals $S_z,S_x,S_y$, and $\sqrt{S_x^2+S_y^2}$, as labeled on each panel (same data as in fig.\ref{fig_3}f). The latter is the calculated radial displacement map of the trapped particle, relative to the center of the scanning trap. 
The first row shows the raw data, while the second row shows the pixel-corrected images, where the transverse displacement of the tip at each pixel is considered to correct the trap transverse position. The images are not high-pass filtered.}
\label{fig_SI_xycorrection}
\end{figure}

An image acquisition starts with the calibration of the particle that will be employed for the scan. This consists initially of a few axial  position-displacement curves on bare glass with the two particle configurations in order to discern the presence and orientation of the tip (briefly blocking the laser between the measurements allows us to obtain both tip up and tip down). Secondly, the traces are recorded with the particle in the liquid bulk to obtain the power spectral densities (PSD) of the signals (and optionally the correlations and Allan variance), which are fit by Lorentzian functions. The position-displacement curve is used to quantify the $z$-sensitivity ($\beta_z$), while the fit of the PSD of $S_z$ provides the other two independent measurements to obtain the stiffness $k_z$ and drag $\gamma_z$. 
For the transverse directions $x$ and $y$, we rely on the theoretical value of the drag of a cylinder $\gamma_\perp$, setting $\gamma_x = \gamma_y = \gamma_\perp$. The PSDs of the signals $S_x, S_y$  then provide the stiffness $k_x, k_y$ and sensitivity $\beta_x, \beta_y$.
Once the drag, stiffness, and sensitivity are obtained for the three directions, the signals $S_z,S_x,S_y$ can be mapped to meters and the force can be quantified \cite{berg2004power}. 

To produce an image, we first define the movement of the sample by the piezo stage (we mainly use 100 steps of 10 or 20 nm in both $x$ and $y$, while $z$ remains constant) and start the scan with the particle in contact to the sample surface.
During the scan, we further acquire from the piezo-driver three voltage signals, which can be converted to nm, and which provide the relative position $x_p(t),y_p(t),z_p(t)$ between the moving sample and the fixed optical trap at each time $t$ ($z_p(t)$ remains constant during the scan). 
The image of the axial displacement of the particle in the trap (signal $S_z$) can then be formed by the set of points ($x_p(t),y_p(t),S_z(t)$), while the images of the transverse displacements ($S_x,S_y$) can be formed by the set of points ($x_p(t),y_p(t),S_x(t)$) and ($x_p(t),y_p(t),S_y(t)$) (see SI fig.\ref{fig_SI_xycorrection}). The points $x_p(t), y_p(t)$ are arranged on a square grid defined by the piezo stage, and can therefore be easily pixelized (each pixel having a surface of 10x10 or 20x20 nm$^2$). This means that all the $x_p(t), y_p(t)$ points belonging to the same pixel are clustered together and the average (or median) value of their corresponding signal ($S_z(t), S_x(t), S_y(t)$ or $\sqrt{S_x^2+S_y^2}$) is calculated and used to set the value of the pixel, rendered in a color code.

Once the signals $S_x$ and $S_y$ are calibrated and converted to nm, this information can be used to correct the position $x_p(t),y_p(t)$ of the trap on the sample to reflect the actual $x,y$ position of the tip in the trap, before the pixels are defined. To do so, we simply add the signals $S_x,S_y$ (in nm) to the signals $x_p(t),y_p(t)$ (in nm), and then we pixelize the points as above. Given that $S_x$ and $S_y$ encode maximum transverse displacements of 1 pixel (10-14 nm, see fig.SI\ref{fig_SI_xycorrection}), we find that this correction modifies only slightly the final image. However it is interesting to note how the image of the radial displacement of the tip (signal $\sqrt{S_x^2+S_y^2}$, see last columns of SI fig.\ref{fig_SI_xycorrection}) underlines the contour of each object in the sample.

\newpage
\section{Supplementary figures}

\clearpage

\clearpage

\clearpage
%\pagebreak

\begin{figure}[h]
\centering
\includegraphics[width=0.5\textwidth]{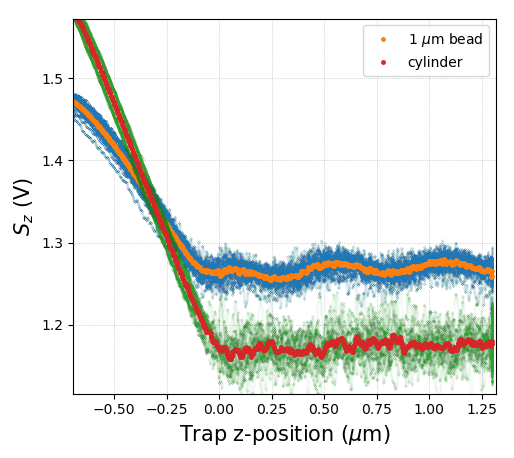}
\caption{Comparison between the position-displacement curves obtained with a 1 $\mu$m polystyrene bead and one cylindrical particle. The oscillations of $S_z$ resolvable with the bead in the trap approaching the glass surface, due to interference with the scattered light \cite{neuman2005measurement}, are not present in the case of our cylindrical particles.}
\label{fig_SI_indent_cyl_bead}
\end{figure}

\clearpage

\clearpage

\clearpage

\clearpage
%\pagebreak

%\newpage
\printbibliography

\end{document}